\documentclass[10pt,letterpaper]{article}
\usepackage{opex3}
\usepackage{geometry}
\usepackage{color,mathptmx,courier,helvet}
\usepackage{graphicx}
\usepackage{amssymb}
\usepackage{amsmath}
\usepackage{epstopdf}
\usepackage{cite}
\usepackage{capt-of}
\usepackage{ulem}

\renewenvironment{abstract}
{\vskip1pc\noindent\begin{center} \begin{minipage}{.8\textwidth} {\bf Abstract: \ } }
{ \\ \vskip-.5pc \noindent \small \copyright \, \number 2015 \hskip.05in
   Optical Society of America \\ \hfil \end{minipage}\end{center}\normalsize\vskip-1.5pc}%
\begin{document}


\title{Strongly squeezed states at 532\,nm based on frequency up-conversion}

\author{Christoph~Baune,$^{1,2}$ Jan~Gniesmer,$^{2,3}$ Axel~Sch\"onbeck,$^{1,2}$  Christina~E.~Vollmer,$^{2}$Jarom\'ir~Fiur\'a\v{s}ek,$^4$ and Roman~Schnabel$^{1,2,\ast}$}
\address{
$^1$Institut f\"ur Laserphysik und Zentrum f\"ur Optische Quantentechnologien, \\Universit\"at Hamburg, Luruper Chaussee 149, 22761 Hamburg, Germany\\
$^2$Institut f\"ur Gravitationsphysik, Leibniz Universit\"at Hannover and Max-Planck-Institut f\"ur Gravitationsphysik (Albert-Einstein-Institut), Callinstrasse 38, 30167 Hannover, Germany\\
$^3$Institut f\"ur Festk\"orperphysik, Leibniz Universit\"at Hannover, \\Appelstrasse 2, 30167 Hannover, Germany \\
$^4$Department of Optics, Palack\'y University, 17. listopadu 12, 77146 Olomouc, Czech Republic
}
\email{$^\ast$\,roman.schnabel@physnet.uni-hamburg.de} 

\begin{abstract}
Quantum metrology utilizes nonclassical states to improve the precision of measurement devices. 
In this context, strongly squeezed vacuum states of light have proven to be a useful resource. 
They are typically produced by spontaneous parametric down-conversion, but have not been generated at shorter wavelengths so far, as suitable nonlinear materials do not exist.
Here, we report on the generation of strongly squeezed vacuum states at 532\,nm with 5.5\,dB noise suppression by means of frequency up-conversion from the telecommunication wavelength of 1550\,nm. 
The up-converted states are employed in a model Mach-Zehnder interferometer to illustrate their use in quantum metrology.
\end{abstract}


\section{Introduction}
The improvement of measurement devices with nonclassical states has been widely explored in many experiments. 
Squeezed vacuum states of light \cite{Vahlbruch2008,Eberle2010,Stefszky2012} in particular have successfully been employed in quantum metrology; they were used to enhance the sensitivity in spectroscopy \cite{Polzik1992}, imaging \cite{Treps2003} and gravitational wave detectors \cite{Schnabel2010,McClelland2011,LSC2011,LSC2013,Grote2013}.
So far, squeezed states were only used to enhance the sensitivities in experiments, where the operating wavelength was in the near-infrared regime.
However, spectroscopic measurements are performed at various optical wavelengths -- including the visible regime.
Consequently, squeezed states at these wavelengths are an interesting resource for nonclassical sensitivity improvements, too.
In addition, the shot-noise sensitivity of interferometric measurement devices generally increases with the inverse of the wavelength giving another motivation to generate squeezed vacuum states at short wavelengths.

Squeezed vacuum states of light are most successfully generated at near-infrared wavelengths by degenerate parametric down-conversion \cite{Mehmet2011} from a strong second-harmonic field. 
However, the generation of squeezed states at visible wavelengths by using parametric down-conversion exhibits technical difficulties.
Intense second harmonic fields at ultraviolet wavelengths would be required, which leads to high absorption and photorefractive damage \cite{Koechner2006}.
Other techniques like four-wave mixing \cite{Slusher1985} also have not been successful to produce strongly squeezed vacuum states at short wavelengths so far.  
Furthermore, self-phase modulation \cite{Bergman1991} and second-harmonic generation \cite{Tsuchida1995} techniques are not able to produce squeezed vacuum states, i.e. squeezed states without a carrier field. 
 
Frequency up-conversion of squeezed states from near-infrared wavelengths into the visible spectrum evades the problem of intense ultraviolet fields. 
The signal combines with a strong idler beam via sum-frequency generation to generate a signal at a higher optical frequency.
The up-converted signal's wavelength is the shortest one involved in the process and crystal damage is avoided because its intensity is very low.
If the frequency conversion is very efficient, the quantum statistics are maintained \cite{Kumar1990,Huang1992}; it was shown that this \textit{quantum} up-conversion maintains features like $g^{(2)}$-functions with values smaller than unity \cite{Rakher2010} or quantum non-Gaussianity \cite{Baune2014, Fiurasek2015}.

The up-conversion of squeezed vacuum states of light from 1550 to 532\,nm was shown for the first time in \cite{Vollmer2014}, where nonclassical noise suppression of 1.5\,dB below shot noise was achieved. 
Here, we present the design of nonlinear cavities optimized for low pump powers and demonstrate significantly higher squeezing in the up-converted field.
The level of squeezing clearly exceeds the 3\,dB threshold, often considered a benchmark level in quantum applications. 
In metrology this value provides a signal to shot noise enhancement that corresponds to doubling the light power when using coherent states only.

\section{Efficient frequency up-conversion of squeezed vacuum states}
	\begin{figure}[htb]
	\centering
	\includegraphics[width=.82\textwidth]{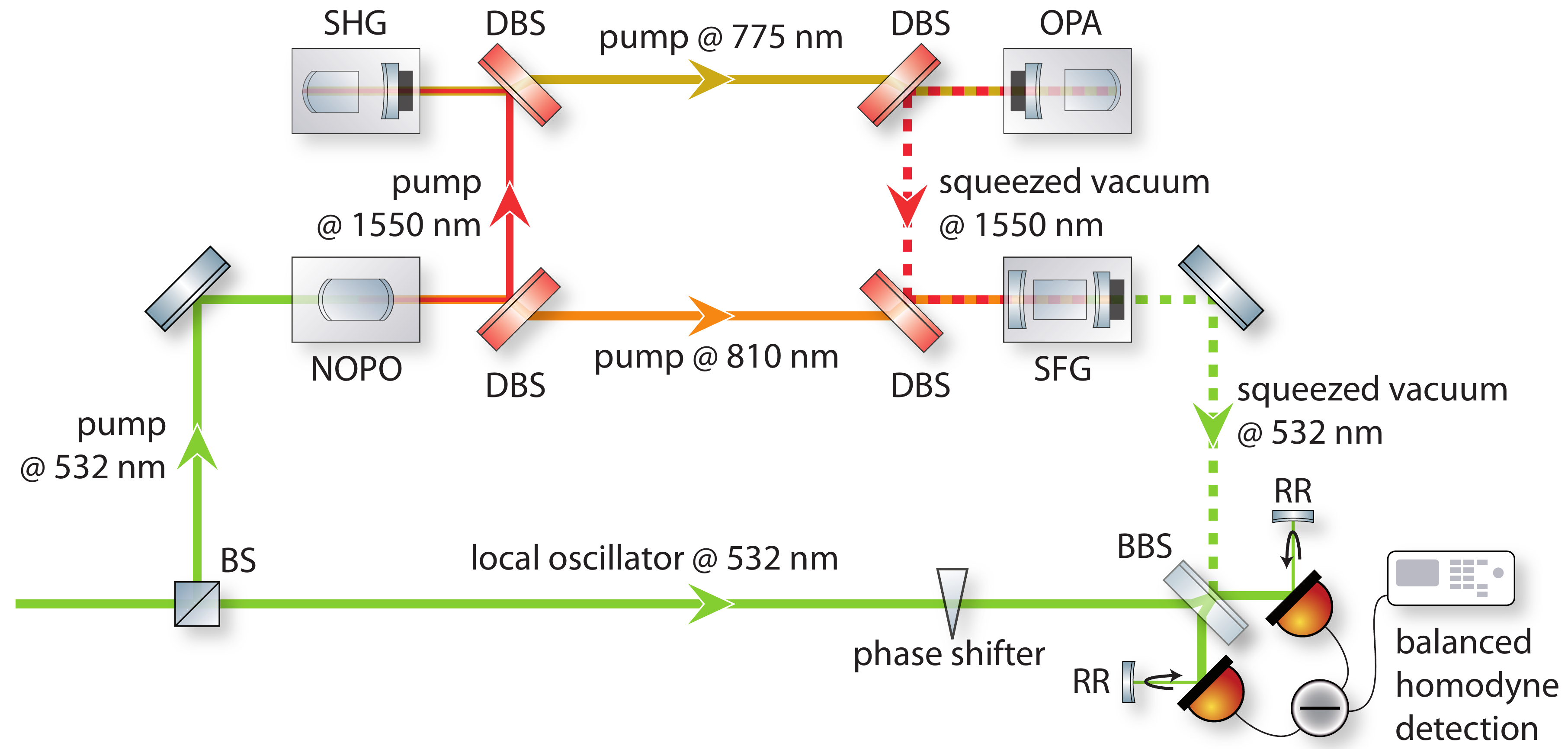}
	\caption{Schematic of the experimental setup involving four nonlinear optical processes. A 532\,nm beam of about 0.7\,W power was split at a low transmission beam splitter (BS). 
	Bright 810 and 1550\,nm fields were produced via cavity-assisted non-degenerate optical parametric oscillation (NOPO). 
	Squeezed vacuum states (illustrated by dashed lines) at 1550\,nm were produced in a degenerate optical parametric amplification cavity (OPA) and up-converted in the sum-frequency generation cavity (SFG) to 532\,nm.
	All cavities are decoupled from each other with optical isolators (not shown).
	The local oscillator for homodyne detection was provided by the second output port of the BS to ensure frequency stability with the up-converted state.
	SHG: second harmonic generation, DBS: dichroic beam splitter, BBS: balanced beam splitter, RR: retro-reflector.}
	\label{fig:SetupSquecov}
	\end{figure}

\begin{figure}[htb]
	\centering
		\includegraphics[width=.99\textwidth]{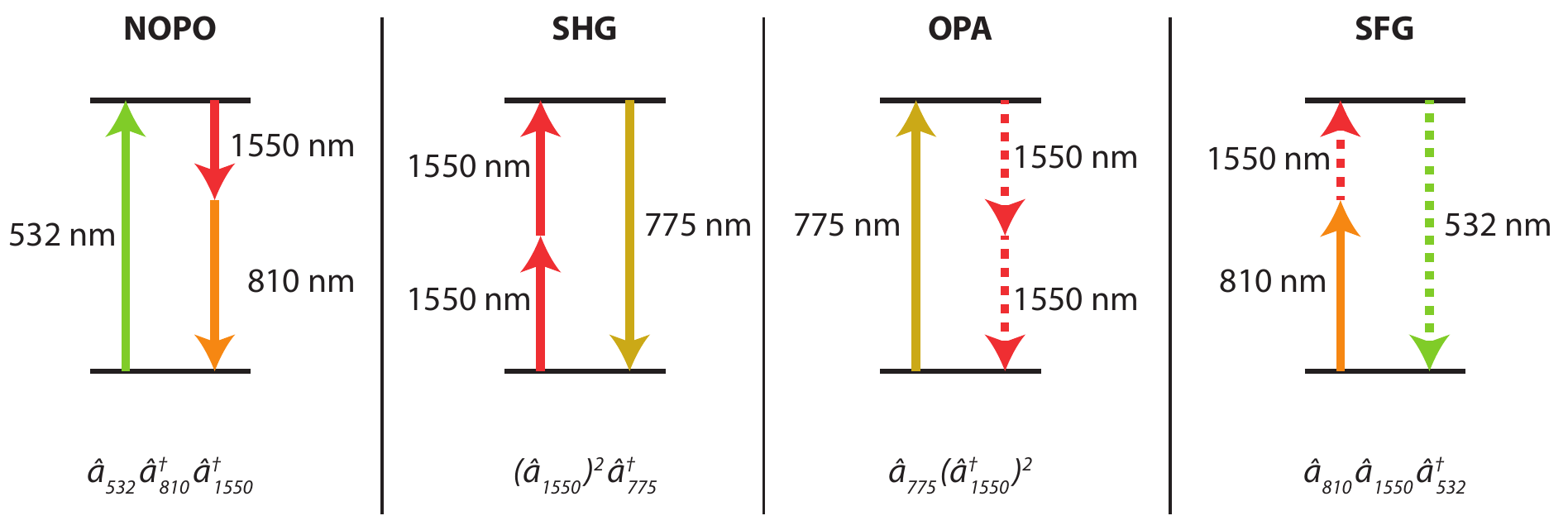}
	\caption{Energy level diagrams of the nonlinear processes involved in the experiment: Non-degenerate optical parametric oscillation (NOPO), second harmonic generation (SHG), optical parametric amplification (OPA) and sum-frequency generation (SFG). Strong classical beams are illustrated by solid lines and weak quantum states by dashes lines. Underneath each picure the operator of the corresponding process is outlined. The coupling constants and the hermitian conjugate parts of full Hamiltonians are omitted for readability.}
	\label{fig:LevelDiagrams}
\end{figure}

The setup for quantum up-conversion of squeezed vacuum states from 1550 to 532\,nm is schematically shown in Fig.~\ref{fig:SetupSquecov}.
A continuous-wave 532\,nm beam (700\,mW) was used to pump a non-degenerate optical parametric oscillator (above threshold, NOPO).
The generated fields at 1550 and 810\,nm (100 and 190 mW) pumped a second harmonic generation cavity (SHG) and the sum-frequency generation cavity (SFG), respectively.
An illustration of all nonlinear processes is shown in Fig.~\ref{fig:LevelDiagrams}.

In contrast to our previous experiment reported in \cite{Vollmer2014}, here, the SHG and the optical parametric amplification cavity (OPA) were optimized for the low pump powers available in our setup.
The nominal reflectivities of the incoupling mirrors were chosen to be 85\,\% at 1550\,nm and 97.5\,\% at 775\,nm for both cavities, while a highly reflective mirror for both wavelengths was directly coated onto the opposite crystals' surfaces.
Simultaneous resonance of the fundamental and harmonic wavelengths was achieved by temperature fine-tuning of the periodically poled potassium titanyl phosphate (PPKTP) while maintaining quasi-phase matching conditions. 
If the doubly resonant condition and quasi-phase matching could not be achieved simultaneously right away, the crystal for the NOPO process was iteratively temperature tuned to slightly change its output wavelengths.

The advantage of a doubly resonant SHG cavity is that the \textit{harmonic} light can be used for Pound-Drever-Hall locking of the cavity length providing an error signal that is not attenuated with increasing conversion efficiency. A conversion efficiency of more than 75\,\% was measured (for a pump power of 100\,mW) -- compared to only 18\,\% in our previous experiment \cite{Vollmer2014}. 

We estimated that more than 10\,dB of squeezing at 1550\,nm was coupled into the SFG. 
Due to the compactness of our setup the fringe visibility at the homodyne detector at 1550\,nm was limited to 95\,\%, and the actual squeezing strength available for up-conversion could not be directly measured.
However, we still measured more than 8.8\,dB nonclassical noise suppression with a respective anti-squeezing of 18\,dB, which indicates the presence of significantly higher squeezing values compared to 4\,dB available for up-conversion in \cite{Vollmer2014}. 

The SFG cavity was modified from what was reported in reference \cite{Vollmer2014} by changing the incoupling mirror reflectivities. 
To lower the effect of intra-cavity loss on the squeezed input field the reflectivity at 1550\,nm was reduced to 91\,\%. Consequently, the pump power build-up had to be enhanced and the reflectivity at 810\,nm was increased to 97\,\%.
The maximum up-conversion efficiency was thereby improved from 84.4\,\% to 90.2\,\%.

The homodyne detector at 532\,nm was built with two photodiodes from Hamamatsu (Type S5973\hbox{-}02).
The protection windows were removed and the diodes were tilted horizontally by about 45\,degrees.
Surface reflections were decreased by using p-polarized light. 
We could not see improvements by increasing the angle even further.
Residual reflections were re-focused onto the photodiodes with highly reflective concave mirrors (focal length 25\,mm). 
The detection efficiency of the photodiodes could thereby be increased from about 85\,\% to up to 90\,\%, which was determined independently by directly measuring the photocurrent when a diode was illuminated with a well known light power.

	\begin{figure}[htbp]
	\begin{minipage}{0.48\textwidth}
	\centering\includegraphics[width=0.99\textwidth]{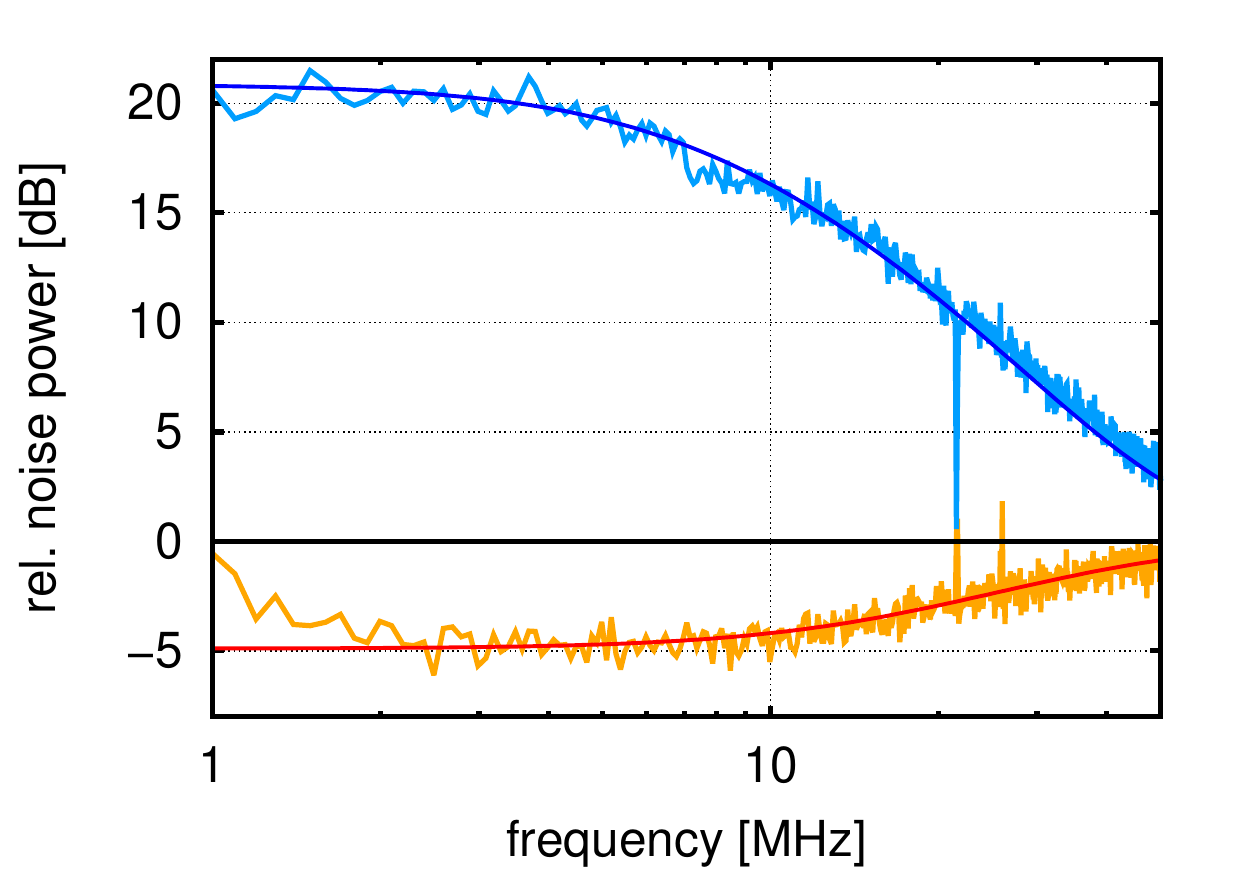}
		\end{minipage}
		\hspace{1mm}\vline\hspace{1mm}
	\begin{minipage}{0.48\textwidth}
	\centering\includegraphics[width=0.99\textwidth]{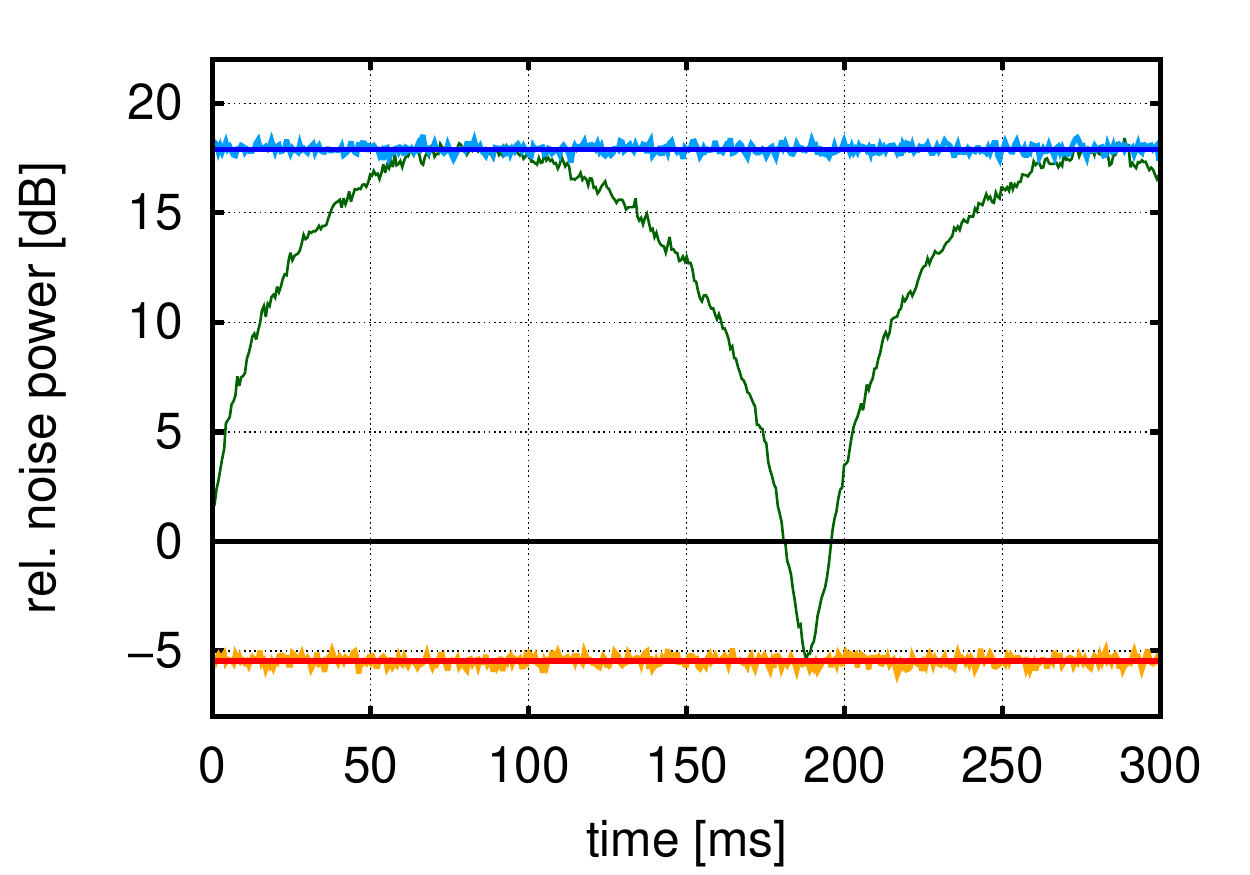}	
	\end{minipage}
	\caption{Measurement of the up-converted squeezed vacuum states at 532\,nm. 
	On the left hand side, we show a power spectrum of the squeezed (orange) and anti-squeezed (cyan) quadrature noise normalized to the vacuum noise reference as well as a theoretical model (red and blue). 
	It was recorded with a resolution bandwidth of 100\,kHz.
	The dark noise was subtracted from the data to allow for a comparison with the theory. 
	For frequencies below 2\,MHz the dark noise of our homodyne detector dominated the measurement and the dark noise correction was too poor to reveal the full squeezing strength.
	The spikes at 21\,MHz and 26\,MHz were due to electronic pick-up in the dark noise caused by a modulation frequency used for cavity locking.
	On the right hand side, our best zero-span measurement at a Fourier frequency of 5\,MHz is shown, recorded with a resolution bandwidth of 300\,kHz and without corrections for dark noise. 
  Here, also a continuous phase scan of the signal quadrature (green) is shown.
	The variance of the squeezed quadrature dropped 5.5\,dB below the shot noise, whereas the anti-squeezed quadrature variance was 18\,dB above.
	The asymmetry is due to the optical losses to which squeezing is more sensitive than anti-squeezing.
	}	
	\label{fig:Squecov2}
	\end{figure}
Figure~\ref{fig:Squecov2} shows a measurement of a power spectrum and a zero-span measurement of the signal field.
We demonstrated a non-classical noise suppression over a wide frequency range from 1 to 50\,MHz.
About 5.5\,dB shot noise reduction was measured at a sideband frequency of 5\,MHz with a respective anti-squeezing of 17.9\,dB.
This state corresponded to an initial squeezing value of 19.3\,dB subject to 27\,\% optical loss.
(To calculate the optical loss, the dark noise was subtracted from the data, which led to slightly higher measured squeezing and anti-squeezing values; -5.55 and 17.94\,dB, respectively.) 
We estimated the contribution of phase noise to be negligible in our experiment. 
The major optical loss sources were the up-conversion efficiency and the quantum efficiency of the photodiodes in the homodyne detector (both about 10\,\% losses). 
Minor contributions came from non-perfect mode-matchings (2.5\,\% loss), a limited visibility at the homodyne detector (2\,\% loss), and propagation losses (5\,\% loss, mainly induced by an optical isolator in the path between the OPA and the SFG).

The theoretical fits for the (anti-)squeezed power spectra $S^-$ ($S^+$) were obtained with the formulae (based on \cite{Collett1984})
\begin{equation}
S^+(\omega)=1+\eta\frac{\kappa^2}{\kappa^2+\omega^2}\frac{4\gamma|\epsilon|}{(\gamma-|\epsilon|)^2+\omega^2},
\end{equation}
\begin{equation}
S^-(\omega)=1-\eta\frac{\kappa^2}{\kappa^2+\omega^2}\frac{4\gamma|\epsilon|}{(\gamma+|\epsilon|)^2+\omega^2},
\end{equation}
where $\omega/2\pi$ is the sideband frequency, $\eta$ is the detection efficiency, $\gamma/2\pi$ and $\kappa/2\pi$ are the bandwidths (HWHM) of the OPA and SFG, respectively, and $\epsilon$ is the pump parameter.
The parameters ($\gamma=2\pi\,$60\,MHz, $\kappa=2\pi\,$40\,MHz, $\epsilon=0.77\,\gamma$) were in good agreement with independently determined values. 
We deduced $\gamma$ from the shape of a measured squeezing spectrum at 1550\,nm. The power dependent parameter $\epsilon$ was measured by comparing the OPA's input power with its measured threshold power. The SFG bandwidth $\kappa$ was determined by scanning the SFG cavity and comparing the Airy pattern with frequency markers from a phase modulation \cite{Brueckner2010}.

\section{Quantum enhancement of a Mach-Zehnder interferometer at 532\,nm}
	\begin{figure}[hbt]
	\begin{minipage}{0.48\textwidth}
	\centering\includegraphics[width=.99\textwidth]{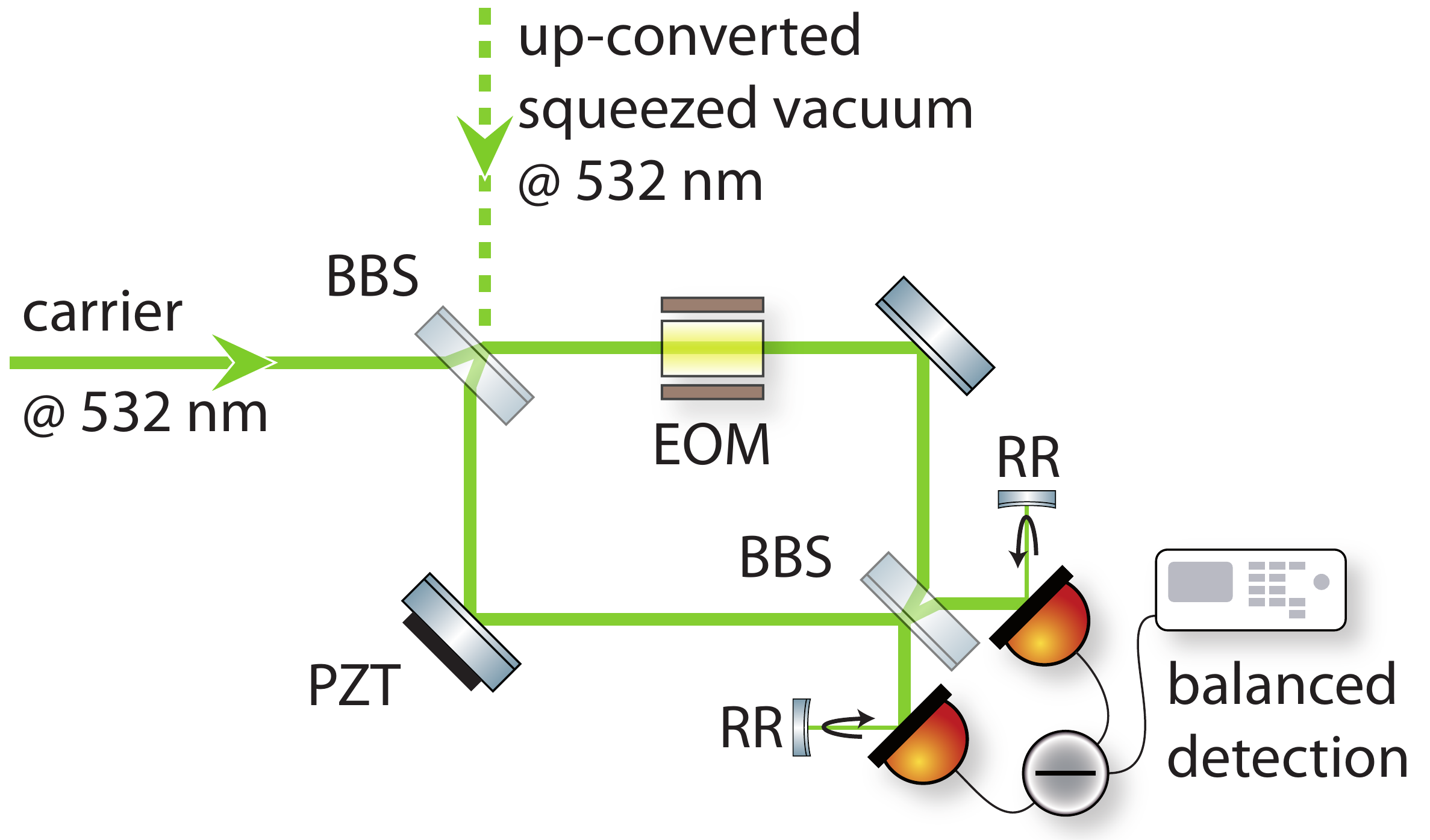}	
	\end{minipage}
	\hspace{1mm}\vline\hspace{1mm}
	\begin{minipage}{0.48\textwidth}
	\centering\includegraphics[width=0.99\textwidth]{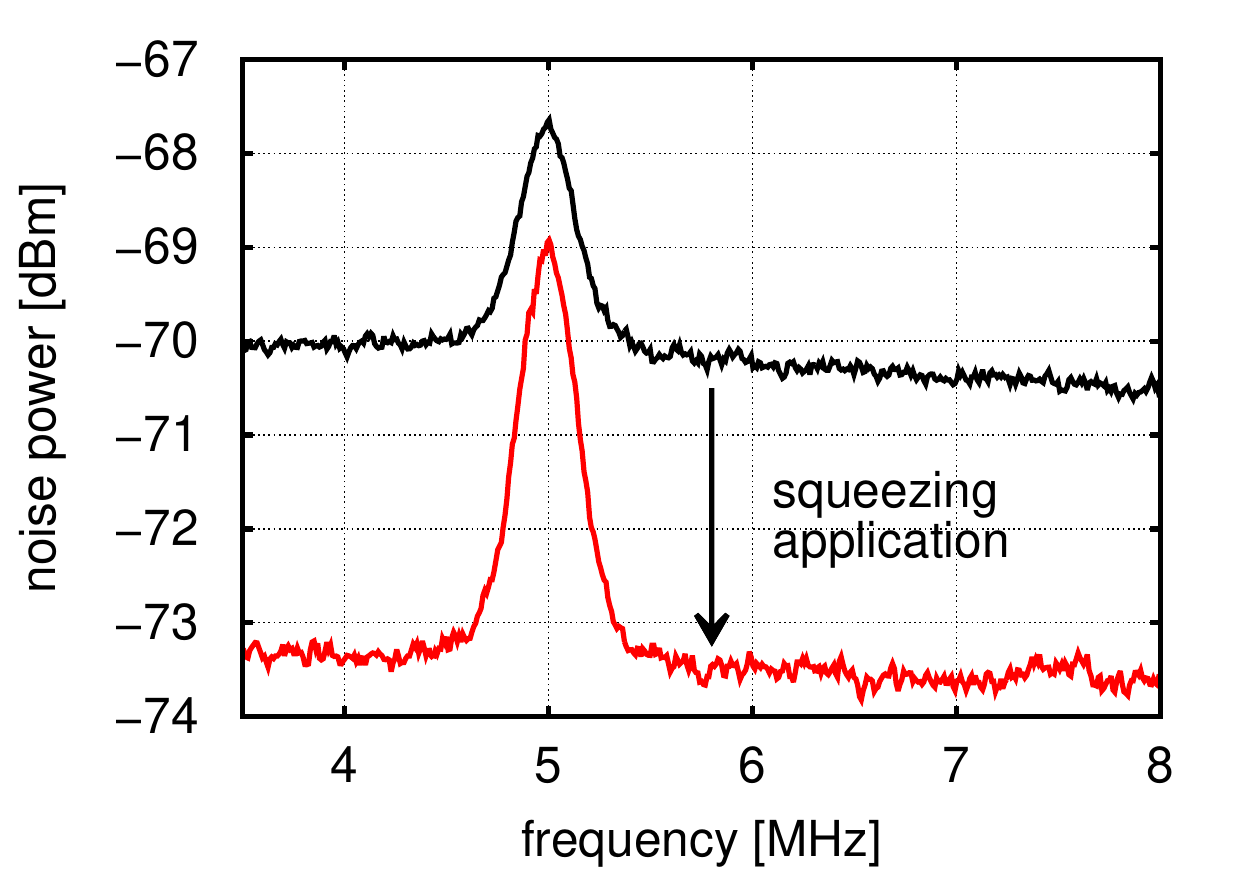}
	\end{minipage}
	\caption{Setup (left) and experimental results (right) of a squeezed light enhanced Mach-Zehnder interferometer. 
	The carrier at 532\,nm and the up-converted squeezed vacuum were the inputs of the interferometer, that consisted of two highly reflective mirrors and two balanced beam splitters (BBS). 
	The two output ports were analyzed with a balanced detector. 
	An electro-optic modulator (EOM) in one of the interferometer arms generated a phase modulation at 5\,MHz, which is clearly visible in the spectrum (right). 
	When the up-converted squeezing was applied, the noise floor dropped and the signal-to-noise ratio improved.
	The resolution and video bandwidth were 300\,kHz and 300\,Hz, respectively, and no dark noise subtraction was performed. 
	PZT: piezo-electric transducer to lock the interferometer at mid-fringe, RR: retro-reflector.}	
	\label{fig:Slemzi}
	\end{figure}
	
	The Mach-Zehnder interferometer consisted of two beam splitters and two highly reflective mirrors. 
A schematic of the device is shown in Fig.~\ref{fig:Slemzi}.	
An electro-optic modulator (EOM) generated a phase modulation at 5\,MHz. 
The output was measured with a balanced detector.
Again, the reflected light was re-focused onto the photodiodes to increase the detection efficiency.
We mounted one of the highly reflective mirrors of the Mach-Zehnder interferometer onto a piezo-electric transducer.
This enabled us to scan the interferometer's phase to adjust the visibility and eventually lock it to mid-fringe.
The signal could be detected with a higher signal-to-noise ratio compared to vacuum (black), when squeezed vacuum (red trace) was injected into the signal port of the interferometer. 
The observation of about -3.3\,dB squeezing improves the phase sensitivity by the equivalent of a 2.1-fold increase in coherent light power, i.e. by a factor of $\sqrt{2.1}=1.46$.
A lower conversion efficiency (80\,\%) and additional optics (including the EOM) increased the optical loss, which degraded the observable squeezing from 5.5\,dB to 3.3\,dB.
The lower conversion efficiency was mainly caused by a degraded mode shape of the 810\,nm pump beam, possibly caused by a dust particle on the NOPO crystal's surface to which the access is difficult.
However, this result reaches the regime of practical usefulness in shot noise limited experiments. The measured quantum enhancement is comparable to large state-of-the-art squeezed light interferometers at infrared wavelengths like GEO\,600 (despite their much higher complexity)\cite{Grote2013}.

\section{Conclusion and outlook}
We demonstrated frequency up-conversion of strongly squeezed vacuum states of light with a conversion efficiency of up to 90.2\,\%. 
The up-converted states at a wavelength of 532\,nm were measured with a noise suppression of up to 5.5\,dB below shot noise, when the alignment, servo phase control as well as the temparature for phase matching of all subsystems -- which included four nonlinear cavities -- were optimized.
A squeezing resonator that was also resonant for the pump light (doubly resonant) helped to produce good squeezing levels, even for the low pump power available.
The linewidth of the up-conversion cavity was increased to minimize the effect of intra-cavity loss for given pump powers \cite{Samblowski2014}.
Furthermore, the detection efficiency of the homodyne detector was optimized by re-focussing reflected light onto the chip.
The observed squeezing spectrum is in excellent agreement with our theoretical model that incorporates transfer functions of not only the squeezing resonator but also the up-conversion cavity.

The compatibility of the up-converted states with an application in a Mach-Zehnder interferometer was demonstrated by a nonclassical improvement of the signal-to-noise ratio by a factor of 1.46.
This result illustrates that highly efficient up-conversion of squeezed vacuum states opens the possibility of sensitivity enhancements of phase measurements, such as those for gravitational wave detectors, where the operating wavelength is reduced to 532\,nm \cite{Decigo2006}.
Furthermore, spectroscopic measurements might greatly benefit from this technique, as in principle the entire visible wavelength regime can be covered by means of quantum up-conversion.

The optimum working point of our experiment was difficult to maintain, as infrequent mode hops of the NOPO slightly changed the precise wavelengths of the pump fields at 810 and 1550\,nm \cite{Eckardt1991,Samblowski2012}. 
Consequently, the phase matching conditions of the subsequent nonlinear cavities varied and the conversion efficiencies decreased.
This problem could possibly be circumvented by amplifying the 1550\,nm pump field for the SHG with a fiber amplifier to ensure that enough pump power for the OPA is always available to produce strongly squeezed states.
In this case, only the NOPO and SFG have to be adjusted to provide high up-conversion efficiency.

\section*{Acknowledgments}
This work was supported by the Deutsche Forschungsgemeinschaft (DFG), Project No. SCHN~757/4-1, by the Centre for Quantum Engineering and Space-Time Research (QUEST), and by the International Max Planck Research School for Gravitational Wave Astronomy (IMPRS-GW). 
J.F. acknowledges financial support from the EU FP7 under Grant Agreement No. 308803 (project BRISQ2) cofinanced by MSMT CR (7E13032).

\end{document}